\documentstyle[12pt]{article}
\footskip- 15mm
\begin{document}
\begin{center}
A SIMPLE METHOD OF CHAOS CONTROL\\
SHAHVERDIEV E. M.,\footnote{e-mail:shahverdiev@lan.ab.az}\\
Institute of Physics, 370143 Baku,Azerbaijan\\
\end{center}
It is well-known that some dynamical systems depending on the value 
of systems'parameters exhibit  unpredictable,chaotic behaviour[1-
6].Such a situation makes impossible long-range prediction of system's 
behavior, but paradoxically allow one to control this behavior with 
tiny perturbations (see,e.g. [7-12] and references therein). The 
seminal papers [7-8] induced avalanche of research works in the 
theory of control of chaos in synergetics.Chaos synchronization in
dynamical systems is one of such ways of controlling chaos. According 
to [7-8] synchronization of two systems occurs when the trajectories 
of one of the systems will converge to the same values as the other 
and they will remain in step with each other. For the chaotic systems 
synchronization is performed by the linking of chaotic systems with a 
common signal or signals (the so-called drivers): suppose that we have 
a chaotic dynamical system of three or more state variables
(it is well-known that for chaotic behaviour in continous dynamical 
systems thenumber of state variables should be no smaller than three 
[3-4]). According to [7-8] in the above mentioned way of chaos 
control one or some of these state variables can be used as an input 
to drive a subsystem consisting of remaining state variables and which 
is a replica of part of the original system.In [7-8] it has been shown 
that if all the Lyapunov exponents  (or the largest Lyapunov 
exponent) or the real parts of these exponents for the subsystem are 
negative then the subsystem synchronizes to the chaotic evolution of 
original system.If the largest subsystem Lyapunov exponent is not 
negative then as it has been proved in [13] synchronism is also 
possible if a nonreplica system constructed according some rule is 
used instead of replica system. The interest to the chaos 
synchronization in part is due to the application of this phenomenen 
in secure communications, in modeling of brain activity  and 
recognition processes,etc [7-12]. Also it should be mentioned that 
this method of chaos control may result in the improved performance 
of chaotic systems (see e.g.[12] and references therein).As it has 
been shown in [13] from the application viewpoint using of nonreplica 
systems has some advantages over the replica approach to the
chaos synchronization. The above-mentioned chaos synchronization 
method [7-8] (replica approach) is applied to different chaotic 
dynamical systems [7-12]. As it is already underlined recently a new 
approach-nonreplica approach to chaos synchronization is proposed in 
[13].A detailed analysis of this paper shows that for high dimensional 
systems the calculation of Lyapunov exponents in general 
requires to solve high order algebraic equations or to recourse to 
the help of numerical simulations.\\
This paper is dedicated to the chaos synchronization in some 
dynamical systems of N-dimensionality within the nonreplica 
approach.In this report a simple method to make all the Lyapunov 
exponents negative is proposed.This is the main feature of my paper.\\
Suppose that an autonomous dynamical system under study has N state 
variables:
$$\frac{dx_{1}}{dt}=f_{1}(x_1, x_{2}, \cdots, x_{N},a_{1}, a_{2}, 
\cdots, a_{N}),$$
$$\frac{dx_{2}}{dt}=f_{2}(x_1, x_{2}, \cdots, x_{N},a_{1}, a_{2}, 
\cdots, a_{N}),$$
$$\vdots\hspace*{14cm}(1)$$
$$\frac{dx_{N}}{dt}=f_{N}(x_1, x_{2}, \cdots, x_{N},a_{1}, a_{2}, 
\cdots, a_{N}),$$
where $x_{1}, x_{2},\cdots,x_{N}$ are state variables,$f_{1}, f_{2}, 
\cdots, f_{N}$
are sufficiently smooth functions of $x_{1}, x_{2}, \cdots, x_{N}$
and $a_{1}, a_{2},\cdots, a_{N}$,and $a_{1}, a_{2}, \cdots, a_{N}$ are
parameters.
Let
$$x_{1}^{ss}, x_{2}^{ss},\cdots, x_{N}^{ss},\hspace*{10cm}(2)$$
be the steady state solutions (fixed points) to the original 
nonlinear dynamical system (1).Also suppose that for some values of 
parameters the system (1) behaves chaotically. As it is known from  
[13] while performing chaos synchronization within replica approach 
one  deals with the response system whose dimensionality is less than 
the dimensionality of the original nonlinear system.But it is trivial 
that for high dimensional original nonlinear system even in the case 
of replica approach  response system's dimensionality could be
high.Also it is well-known that within nonreplica approach response 
system's dimensionality is equal to the dimensionality of the 
original nonlinear system.
That is why without loss of generality I will investigate the case of 
nonreplica approach in order to deal with highest possible 
dimensionality.As it was already mentioned above,the possibility of 
chaos synchronization essentially depends on the sign of the Lyapunov 
exponents.To be more precise, these exponents should be negative.\\
According to [13], within nonreplica approach the response system 
contains some arbitrary constants added according to some rule.The 
presence of these arbitrary constants allows one to be more flexible 
to achieve chaos synchronization.\\
Without loss of generality, take the state variable $x_{1}$ as a 
driver.Then using approach developed in [13] I construct the 
following nonreplica response system(with the superscript"nr"):
$$\frac{dx_{1}^{nr}}{dt}=f_{1}(x_1, x_{2}^{nr}, \cdots, 
x_{N}^{nr},a_{1}, a_{2}, \cdots, a_{N})+\alpha_{1}(x_{1}^{nr}-
x_{1})=F_{1},$$
$$\frac{dx_{2}^{nr}}{dt}=f_{2}(x_1, x_{2}^{nr}, \cdots, 
x_{N}^{nr},a_{1}, a_{2}, \cdots, a_{N})+\alpha_{2}(x_{1}^{nr}-
x_{1})=F_{2},$$
$$\vdots\hspace*{14cm}(3)$$
$$\frac{dx_{N}^{nr}}{dt}=f_{N}(x_1, x_{2}^{nr}, \cdots, 
x_{N}^{nr},a_{1}, a_{2}, \cdots, a_{N})+\alpha_{N}(x_{1}^{nr}-
x_{1})=F_{N},$$
(Here it is necessary to underline the following point: in order to 
construct the response system in fact I added to the  right-hand side 
of the initial nonlinear equations  linear terms on the difference of 
original and response system's variable. It is made, as my task is to 
demonstrate the simplest way of achiving chaos synchronization: 
according to [13], in principle  one can construct the response 
system by adding to the original nonlinear system arbitrary 
functions, which vanish when chaos synchronization is achieved.)
The eigenvalues of the Jacobian matrix of the nonreplica system
$$J=\frac {\partial (F_{1}, \cdots, F_{N})}{\partial (x_{1}^{nr}, 
\cdots, x_{N}^{nr})}, \hspace*{5cm} (4)$$
satisfies the following equation:
$$\lambda^{N}+p_{1}\lambda^{N-1}+p_{2}\lambda^{N-
2}+\cdots+p_{N}=0,\hspace*{2cm}(5)$$
where $p_{1}, p_{2}, \cdots,p_{N}$ are in general the functions of the
arbitrary constants $\alpha_{1}, \alpha_{2},\cdots, \alpha_{N}$, 
parameters
$a_{1}, a_{2}, \cdots, a_{N}$, and solutions of the original 
nonlinear system
(1) $x_{1}(t), x_{2}(t), \cdots, x_{N}(t)$.It is well-known that in 
general case
with some exceptions it is highly problematic to find the exact 
analytical solution of the system of nonlinear equations.This fact 
creates immense difficulties in the treatment of equation (5) 
analitically. \\
As it was mentioned above the task of this paper is to make negative 
all the roots of equation (5) without the need of performing tedious 
numerical and analitycal calculations. As the analyses show there are 
some class of dynamical systems, which could be explored from this 
point of view. In other words, albeit in general the coefficients 
$p_{1}, p_{2}, \cdots,p_{N}$ are the functions of 
time, in some cases the equation (5) could be solved easily .\\
For example, it is trivial, that in the case of constant Jacobians 
[13] of the initial nonlinear system these coefficients are time-
independent (below as an example one of the R\"ossler models is 
investigated), which allows to treat equation (5) quite easily.But 
there is a wide class of dynamical systems, chaos synchronization in 
which can be treated with relative ease even in more general
case.I mean dynamical systems with bounded solutions.It is widely 
known that many dynamical systems with dissipative nature have bounded 
solutions in the sense that solutions of these systems never goes to 
infinity. It is well-known that the classical Lorenz system is one of 
well-studied dissipative dynamical systems with bounded solitions 
(see, e.g.[1, 2, 4-5]).\\
Below as an example I will investigate this classical Lorenz model in 
the relatively unexplored case.\\
But first I present the more general approach developed for the 
bounded systems.So, suppose that the original nonlinear system has  
bounded solutions.As it has been shown by E.N.Lorenz in [14], the 
dissipative systems of the form 
$$\frac{dx_{i}}{dt}=\sum_{j, k=1}^{N}a_{ijk}x_{j}x_{k}-
\sum_{j=1}^{N}b_{ij}x_{j}+c_{i},\hspace*{3cm}(6)$$
with the constants chosen so that $\sum a_{ijk}x_{i}x_{j}x_{k}$ 
vanishesidentically and $\sum b_{ij}x_{i}x_{j}$ is positive definite,
have bounded solutions.Using the boundedness of the solutions and 
more crucially the presence of arbitrary constants one can try to make 
the roots of equation (5) negative without conducting explicit 
calculations of these roots.  \\
Thus, in general I obtain the equation of N-dimensionality for 
Lyapunov exponents with coefficients depending on arbitrary 
constants $\alpha_{1}, \alpha_{2}, \cdots, \alpha_{N}$.
Due to the flexibility in choosing the form of nonreplica response 
system, one will be able to obtain eigenvalue equation (5) with 
coefficients containing only linear terms on these arbitrary 
constants.(As it will be clear below  from the investigation of one of 
R\"ossler models, in the case of such linearity in some
cases one can first virtually "choose" any desired negative values 
for Lyapunov exponents and after that calculate "the right" arbtitrary 
constants to achive the necessary goal even more easily.)   \\
For this purpose I choose such a nonreplica response system which 
gives rise to the Jacobian containing all the arbitrary constants 
along one column.It should be noted that if chaos synchronization is 
investigated within  nonreplica approach and the number of driving 
variables more than unity then it is possible to obtain algebraic 
equation of N-th order with coefficients containing also nonlinear 
terms on the arbitrary constants. One should keep in
mind, as a rule the more the number of arbitrary constants,the easier 
to achive our goal of negative Lyapunov exponents. But without loss of 
generality and for the sake of simplicity a case of coefficients with 
linear terms on the arbitrary constants will be studied.In the case of 
nonlinear terms on the arbitrary constants again due to the 
flexibility warranted by the form of the nonreplica
response system it is possible to choose some of these constants so 
that coefficients before $\lambda$'s could contain only linear terms 
on the arbitrary constants.\\
So I have some N order algebraic equation.Suppose that $\lambda_{i}$
($i=1, 2, 3, \cdots, N$) are roots of this equation.It means that the 
equation (5) for the eigenvalues of the Jacobian matrix of the 
nonreplica response system could be presented in the following form:
$$\prod_{i=1}^{N}(\lambda-\lambda_{i})=0 \hspace*{5cm}(7),$$
or\\
$$\lambda^{N}+s_{1}\lambda^{N-1}+s_{2}\lambda^{N-2}+\cdots+s_{N}=0 
\hspace*{2cm}(8),$$
where $s_{1}, s_{2}, \cdots, s_{N}$ are functions of 
$\lambda_{1},\lambda_{2},\dots,\lambda_{N}$:
$$s_{1}=(-1)^{1}\sum_{i=1}^{N}\lambda_{i},$$
$$s_{2}=(-1)^{2}\sum_{i=1}^{N}\sum_{j, 
j>i}^{N}\lambda_{i}\lambda_{j}\hspace*{4cm}(9),$$
$$s_{3}=(-1)^{3}\sum_{i=1}^{N}\sum_{j, j>i}^{N}\sum_{k, 
k>j}^{N}\lambda_{i}\lambda_{j}\lambda_{k},$$
$$\vdots$$
$$s_{N}=(-1)^{N}\prod_{i=1}^{N}\lambda_{i},$$
Now one has a characteristic equation expressed in two ways:1) 
equation (8);2) equation (5) obtained from the calculation of 
eigenvalues of Jacobian matrix of the nonreplica response 
system.Comparing terms with the same order of $\lambda$ it is possible 
to express arbitrary constants in the nonreplica  response system 
through the solutions of the characteristic equation (or vice
versa):
$$p_{1}=s_{1}, p_{2}=s_{2},\cdots,p_{N}=s_{N},\hspace*{6cm}(10)$$
In the equation (10) by replacing the dynamical variables $x_{1}, 
x_{2},\cdots, x_{N}$ by some numbers (for the given value of system's 
parameters) from within the allowable diapason of values of dynamical 
variables one obtains the time-independent $p_{1}, p_{2}, \cdots, 
p_{N}$.(It would be quite reasonable to study the behavior of the 
characteristic equations' coefficients as a function of bounding 
(limiting) values for the original nonlinear system; but
again the free choice of arbitrary constants in the nonreplica 
approach allow one effectively to achive the goal even without such an 
investigation).It is well-known that the necessary and sufficient 
conditions for the roots of eqs.(8) or (5) to have negative real 
parts) are the Routh-Hurwitz criteria. (Below upon investigating the 
examples these conditions will be written explicitly.)
As it will be seen from the represented below examples one could 
quite easily" pick up" the appropriate value and sign for the 
arbitrary constants in the nonreplica approach to make negative the 
real parts of the Lyapunov exponents.\\
Thus the feature of my approach to the chaos synchronization is that 
for some dynamical systems the possibility of chaos synchronization 
could be judged without calculating Lyapunov exponents explicitly.
This feature of approach could be useful from the application point of
view in the sense that the feasibility of synchronization could be 
established
with relative easiness.\\
Now as the first example consider the following nonlinear chaotical 
dynamical system.The system under consideration is of the form 
([15],the fourth model
proposed by R\"ossler in 1977 the so-called model 1977-1V:
$$ \frac {dx}{dt} = -y-z, $$
$$ \frac {dy}{dt} = x,\hspace*{4 cm} (11) $$
$$ \frac {dz}{dt}= a (1-x^2)-bz, $$
According to [15] the dynamical system for  values of parameters $ 
a=0.275, b=0.2 $ (see [16]) exhibits chaotic behaviour. The system 
(10) has the following fixed point:
$$ x=0, z=ab^-1, y=-ab^{-1}, \hspace*{3cm} (12)  $$
The system (11) has three dynamical variables, there is only one 
nonlinear term of a single variable,namely $x$.
I will consider the case, when $x$ variable is the driver.According 
to [13] the following form of nonreplica system (with the subscript 
"nr") is adequate:
$$ \frac {dx_{nr}}{dt}=-y_{nr}-z_{nr}+ \alpha_{1} (x_{nr}-x), $$
$$ \frac {dy_{nr}}{dt}=x+\alpha_{2} (x_{nr}-x) ,\hspace*{3cm} (13) $$
$$ \frac {dz_{nr}}{dt}= a(1-x^{2})-bz_{nr}+\alpha_{3} (x_{nr} -x), $$
As the calculations show the eigenvalues of the Jacobian matrix of 
the system (13) satisfies the following equation:
$$\lambda^{3}+\lambda^{2}(b-\alpha_{1})+\lambda(\alpha_{2}$$
$$+\alpha_{3}- b\alpha_{1})+b\alpha_{2}=0, \hspace*{2.5cm} (14)$$
Suppose that $\lambda_{1}, \lambda_{2}, \lambda_{3}$ are roots of 
this equation.Then using the aboveproposed method (comparing the 
equations (5) and (7)) it is very easy to establish the following 
relationship between the coefficients of eq.(14) and these roots:
$$p_{1}=b-\alpha_{1}=-(\lambda_{1}+\lambda_{2}+\lambda_{3})=s_{1},$$
$$p_{2}=\alpha_{2}+\alpha_{3}-b\alpha_{1}$$
$$=\lambda_{1}\lambda_{2}+\lambda_{1}\lambda_{3}$$
$$+\lambda_{2}\lambda_{3}=s_{2},$$
$$p_{3}= b\alpha_{2}$$
$$=-\lambda_{1}\lambda_{2}\lambda_{3}=s_{3},\hspace*{3cm} (15) $$
As $\lambda_{1},\lambda_{2},\lambda_{3}$ should be negative,
I obtain the following inequalities from the relationships (14):
$$p_{1}= b-\alpha_{1}>0,$$
$$p_{2}=\alpha_{2}+\alpha_{3}-b\alpha_{1}>0,\hspace*{5cm}(16)$$
$$p_{3}= b\alpha_{2}>0,$$
But one should keep in mind that these conditions are not sufficient 
to have negative roots (or roots with negative real parts).According 
to Routh- Hurwitz criteria, for roots with negative real parts, 
additional to (16) condition is required: Namely, the inequality
$$p_{1}p_{2}-p_{3}=b(\alpha_{3}+\alpha_{1}^2)$$
$$-\alpha_{1}(b^{2}+\alpha_{2}+\alpha_{3})>0,\hspace*{0.5cm}(17)$$
also should take place. (In fact, according to [17 ] the positiveness 
of $p_{1}, p_{3}, p_{1}p_{2}-p_{3}$ is sufficient, as the positiveness 
of $p_{2}$ follows from the previous inequalities).\\
As it can be seen from the relationships (16) and (17), it is quite 
easy to make Lyapunov exponents negative by choosing positive values
for $\alpha_{2}$ and quite large negative values for $\alpha_{1}$.
(Here and below on studying the Lorenz model one should keep in mind
that in practice the wide dynamic range for state variables is 
undesirable and this difficulty can be eliminated by a simple 
transformation of variables, see,e.g. [9].)
As the Jacobian in the case of R\"ossler model is constant, one can 
even first choose any desired values for the Lyapunov exponents, 
after that solve the equation (15) to find arbitrary constants.(For 
non-constant Jacobians it is rather difficult to do, because within my 
approach the exact values for time-dependent solutions of the 
original nonlinear system are not necessarily to be
known; also one should aware of nonmonotonic behavior of these 
solutions.) One can see easily that in the case of linear dependence 
of the coefficients of the characteristic equations (5) or (8) on the 
arbitrary constants, the task is the simplest one. Presenting the 
application of the proposed method one should keep in mind the case of 
this particular R\"ossler model is the trivial one in the sense that 
one deals with the constant Jacobian and therefore the coefficients
before $ \lambda $'s are time-independent.\\
Here I would like to stress the following conclusion which can be 
derived from the results of the application of the proposed algorithm 
to the R\"ossler model investigated in this report. The studied 
R\"ossler model contains three state variable and only one nonlinear 
term of a single variable $x$.Considering this variable as a driver 
one obtains in essence linear response system,which contains
three arbitrary constants within nonreplica approach to the chaos 
synchronization and by choosing these arbitrary constants one can make 
all the Lyapunov exponents negative and perform synchronization.\\
Using the algorithm proposed in this report it is easy to arrive at 
the same conclusion in the general case:Namely if one has a nonlinear 
dynamical system with an N-dimensional phase space, and if all the 
nonlinear terms are functions of a single variable $ x $, then it is 
always possible to find an N- dimensional linear response system, with 
N arbitrary constants, which will synchronize when driven by $x$, if 
the N constants are adjusted to make all eigenvalues of the constant 
Jacobian matrix negative.The linearity of the response system is 
highly important in the communications applications from the point of 
view of exact recovery of transmitted signals(see [18] and references 
cited therein).
Speaking about the communications applications of the chaos 
synchronization one should also mention that by choosing the arbitrary 
constants one can make all the Lyapunov exponents not only negative, 
but also larger in magnitude.This fact also is very important from the 
application viewpoint. Because, the time required for synchronization 
to take place depends on the value of the largest Lyapunov exponent.\\
Now as the second example of application of the proposed method 
consider the nontrivial case of classical Lorenz dynamical system:
$$\frac{dx}{dt}=\sigma (y-x), $$
$$\frac{dy}{dt}=rx-y-xz, \hspace*{0.7cm} (18) $$
$$\frac{dz}{dt}=xy-bz, $$
It is well-known that the dynamical system (18) for some values of 
parameters exhibits chaotic behaviour [1-10].
The adopted values of parameters followed  by Lorenz and most other 
investigators are: $ \sigma =10 $ and $ b=\frac {8}{3} $.As for
the values of $ r $ for the chaotic behaviour to occur, according to 
the linear stability analysis, for the given values of other 
parameters $ r $ must be larger than critical Rayleigh number 
$ r_{cr} $, see, e.g.[1-3]. At $ r > r_{cr} $ the fixed points of the 
Lorenz system
$$ x_{ss}=y_{ss}=\pm (b(r-1))^{\frac{1}{2}},$$
$$ z_{ss}=r-1,\hspace*{2cm}(19) $$ 
become unstable,and there is a strange attractor over which a chaotic 
motion takes place.\\ 
It is well known that for $\sigma =10 $ and $ b=\frac {8}{3}$ the 
critical Rayleigh number is equal to $ r_{cr}=24.74 $. In [8], while 
investigating the chaos synchronization in Lorenz model the value of 
$ r=60 $ was used. As it was mentioned above, the Lorenz model is a 
classical example of chaotic behavior in low dimensional nonlinear 
dynamical systems, and is one of well studied nonlinear systems. 
Although, the chaos synchronization phenomenen in Lorenz system is 
also investigated in detail, nevertheless there is some gap in the 
study of this phenomenon. Namely, the possibility of chaos 
synchronization in the case of $ z $ variable as a driver has not 
been analyzed thoroughly yet.(To my knowledge, there is only one 
recent paper [19] addressing this issue.In that paper,the chaos 
sinchronization in the case of $ z $ driving is  achieved by 
considering perturbations of the nonlinear system's parameter,
to be more specific the perturbation of the parameter $ r $ was 
considered.In this paper I demonstrate that such a synchronization is 
possible even without parameter perturbations within non-replica 
approach.)\\
As it was shown in [8], in the case of $ z $ variable as a driver 
synchronization of the response subsystem (x,y) with the original 
Lorenz system does not occur for the values of parameters 
$\sigma =10, b=\frac{8}{3}, r=60 $, as one of the sub-Lyapunov 
exponents is positive. Here I will apply the proposed method of
chaos synchronization to this case.\\
Thus, consider the $z$ variable as a driver.Then according to [13], 
in the case of failure of replica approach the following nonreplica 
system (with the subscript "nr") can be used for synchronization 
purposes.     
$$\frac{dx_{nr}}{dt}=-\sigma x_{nr}+\sigma y_{nr}$$
$$+ \alpha_{1}(z_{nr}-z),$$
$$\frac{dy_{nr}}{dt}=rx_{nr}-y_{nr}-x_{nr}z$$
$$+\alpha_{2}(z_{nr}-z) ,\hspace*{0.7cm} (20)$$
$$\frac{dz_{nr}}{dt}= -bz+ x_{nr}y_{nr}+\alpha_{3}(z_{nr} -z),$$
As the calculations show the eigenvalues of the Jacobian matrix of 
the system (20) satisfy the following equation:
$$\lambda^{3}+\lambda^{2}(\sigma+1-\alpha_{3})$$
$$-\lambda (y\alpha_{1}+x\alpha_{2}+(\sigma +1)\alpha_{3})$$
$$+\sigma (r-z)-\sigma)$$
$$-y(\sigma \alpha_{2}+\alpha_{1})-\sigma\alpha_{3}$$
$$-x\alpha_{2}\sigma -(r-z)(x\alpha_{1}$$
$$-\alpha_{3}\sigma)=0, \hspace*{0.3cm} (21)$$
Here $x (t), y(t), z(t)$ are the solutions of the Lorenz system 
(18).\\
According to Routh-Hurwitz criteria, the  sufficient and necessary 
conditions to have roots  with negative real parts  for the equation 
(21) can be written as:
$$\sigma +1-\alpha_{3}>0,$$
$$-y(\sigma \alpha_{2}+\alpha_{1})-\sigma\alpha_{3}$$
$$-\alpha_{2}\sigma -(r-z)(x\alpha_{1}$$
$$-\alpha_{3}\sigma)>0, \hspace*{0.3cm} (22)$$
$$\sigma (\sigma +1)(1-(r-z))+\alpha_{3}^{2}(\sigma +1)$$
$$-(\sigma +1)^{2}\alpha_{3}$$
$$+\sigma y(\alpha_{2}-\alpha_{1})+x(\alpha_{1} (r-z)-\alpha_{2}) $$
$$+\alpha_{3}(y\alpha_{1} +x\alpha_{2})>0$$
To move further I use the fact that solutions of the Lorenz system is 
bounded.The bounding value depends on the relationships between the 
system's parameters and the expression for it could be found in 
different textbooks and papers, see, e.g.[2, 5, 20].
As the solutions of the initial Lorenz model are bounded and one can 
choose the magnitude of the arbitrary constants arbitrarily large or 
small and the sign negative or positive, then it can be seen easily 
from the equation (22), say, by equalizing $\alpha_{3}$ to a large 
negative value to $- 100$),and by choosing  $\alpha_{1}$ and 
$\alpha_{2}$ approximately equal in magnitude, it is possible to make 
negative the real parts of the Lyapunov exponents. For the obtaining 
of the negative Lyapunov exponents, it would be quite helpful,to take
into account the fact that after transition processes in the long 
time limit for $\sigma>>1$ , $x(t) \approx y(t)$ and $z(t)>0$.For the 
given values of the system's parameters it is relatively easy to 
"predict" the right order of arbitrary constants to obtain Lyapunov 
exponents with negative real parts. Really, taking into account the 
above-mentioned equality of $x(t)$ and $y(t)$, also the positiveness 
of $z(t)$ and writing $z(t)=r-1-\epsilon$, where $\epsilon$ is not 
necessarily a small number, one can obtain the following 
expressions for the coefficents of the characteristic equation:
$a_{1}=\sigma +1-\alpha_{3}, 
a_{2}=-((\sigma +1)\alpha_{3}+y(\alpha_{1}+\alpha_{2})+\epsilon),
a_{3}=\alpha_{3}\sigma\epsilon -2\sigma\alpha_{2}y -y(2+\epsilon)$.
From this expressions one can easily establish that larger negative 
values of $\alpha_{3}$ and large positive (if $y<0$) or negative (if 
$y>0$) values of $\alpha_{2}$ will help to satisfy the conditions of 
negativity of real parts of the roots of characteristic
equation $a_{1}>0, a_{2}>0, a_{3}>0, a_{1}a_{2}-a_{3}>0$. Also 
are appropriate the larger negative values of $\alpha_{3}$ with the 
small and close magnitudes of constants $\alpha_{1}, \alpha_{2}$ from 
the point of view  of obtaining of Lyapunov exponents with negative 
real parts. This "right guess" is  confirmed by the numerical 
simulations. Really, for $\sigma =10, b=\frac{8}{3}, r=60$, taking 
$\alpha_{3}=-100, \alpha_{1}=-1,\alpha_{2}=-1$ from the exact 
solution of the equation for the Lyapunov exponents (the initial 
Lorenz model was solved by the fourth-order Runge-Kutta
model) I found the following values for the Lyapunov exponents:
$\lambda_{1}=-2.575, \lambda_{2}=-11.000, \lambda_{3}=-97.425$.
So, just using the boundedness of the dynamical systems (eq.(6)) and 
applying nonreplica approach I was able to perform chaos 
synchronization in Lorenz model.In difference to the approach 
developed in [19], I did't use the system's parameters perturbation.\\
Speaking about the possibility of replacement of the solutions of the 
original nonlinear system with some constant values for the 
calculation of sub- Lyapunov exponents, I would like to stress that 
such an approach for the first time (to my knowledge) was used by the 
authors of the paper [10]. Namely, as it was proved in [10] 
numerically, while calculating the sub-Lyapunov exponents for
dynamical systems, whose chaotic behavior has arisen out of 
instability of fixed points (steady-state solutions) one can replace 
the solutions of the original nonlinear systems with the steady-state 
solutions. Moreover, for some of these systems (e.g., for Lorenz 
system, some of R\"ossler models) sub-Lyapunov exponents of some of 
the unstable fixed points  appears to govern the locking not only to 
chaotic orbits, but also to the periodic orbits. In other words, the 
sub-Lyapunov exponents for the fixed points and the periodic orbit 
also agree with each other. In the light of these results,it would 
be quite interesting to try to calculate not only sub-Lyapunov 
exponents, but also total Lyapunov exponents.\\
Having this in mind, I also calculated (numerically) the total 
Lyapunov exponents of the equation (18) by replacing the time- 
dependent  solutions of the Lorenz model with the non-trivial steady-
state solutions. As the numerical simulations show in general case 
total Lyapunov exponents for the time-dependent and steady-state 
solutions are different. For example, using the above-mentioned
values of system's parameters $\sigma =10, b=\frac{8}{3}, r=60$ and 
taking instead of time-dependent solutions of the Lorenz model the non-
trivial steady-state solution I obtain the following values for the 
total Lyapunov exponents:
$\lambda_{1}=-0.251, \lambda_{2}=-11.000, \lambda_{3}=-99.75$. As it 
can be seen in general these total Lyapunov exponents for the cases of 
time- dependent and time-independent solutions are different(although 
one can see the sharp difference only between one Lyapunov exponents): 
the two others are in satisfactory agreement; by the way, this 
tendency was  established also for other values of system's 
parameters, even for cases  when one of total Lyapunov exponents 
becomes positive. It appears that the satisfactory 
agreement established between two sub-Lyapunov exponents within the 
replica approach has some memory-retaining influence on the two of the 
three of the total Lyapunov exponents in the case of non-replica 
approach to chaos synchronization;at the same time one should be aware 
of the fact that these two of the total Lyapunov exponents are 
different in magnitude from those sub-Lyapunov exponents within
the replica approach. The fact that the Lyapunov exponents are 
different for the cases of time-dependent and steady state solutions 
to the original Lorenz model could be seen from the following argument 
even without explicit numerical and analytical calculations: really 
for the case of steady state solutions the equation (21) gives the 
following expression:
$$\lambda^{3}+\lambda^{2} (\sigma +1-\alpha_{3})$$
$$-\lambda ((\alpha_{1}+\alpha_{2})x_{ss}+\sigma +1)\alpha_{3})$$
$$-2x_{ss}(\sigma \alpha_{2} +\alpha_{1})=0,\hspace*{0.1cm}(23)$$
It can be seen easily from this equation for $\alpha_{1}=0, 
\alpha_{2}=0$ one of the roots of this equation is equal to zero 
exactly.Putting $\alpha_{1}=0, \alpha_{2}=0$ also in the equation (21) 
will not give the same result. Having in mind the above-mentioned 
satisfactory agreement between the sub-Lyapunov exponents for the 
fixed points and the periodic orbits, one can say that in general the 
total Lyapunov exponents  of the  chaotic orbit and  the periodic 
orbits  also will not coincide with each other. Speaking about the 
different values for the total Lyapunov exponents,one should
also to mention that in some special cases by choosing the 
appropriate arbitrary constants, one can obtain close total Lyapunov 
exponents for the cases of both time-dependent and time-independent 
solutions with high degree of accuracy.\\
Really, taking $\alpha_{1}=-10, \alpha_{2}=0, \alpha_{3}=0$ (with the 
above-mentioned set of system's parameters) I obtain that in the case 
of steady-state solutions the real parts of the two Lyapunov exponents 
are equal to $\lambda_{1}=\lambda_{2}=-0.326$. The third Lyapunov 
exponent equals $\lambda_{3}=-10.348$. The numerical calculation of 
total Lyapunov exponents for the time-dependent solutions gives rise 
to the following values:
$\lambda_{1}=\lambda_{2}=0.320, \lambda_{3}=-10.361$. One can see 
that there is a quite good agreement between two cases.But these cases 
could be called "coincidental closeness" and could be explained by the 
choice of arbitrary constants in the nonreplica approach. One more 
point to underline:judging by the form of nonreplica response system, 
it is clear that, in fact I have used the linear feedback method of 
chaos control.Preserving all the arbitrary  constants, one actually 
makes the task of making Lyapunov exponents negative easier. It is 
quite interesting to study this problem in more particular cases, say 
nullifying one or more of these arbitrary constants, in other words 
feedback scheme works only for part of the state variables.By 
studying these cases, I found that to make all Lyapunov exponents 
negative is problematic, even in some cases virtually impossible.For 
example, taking in equation (21) $\alpha_{1}=0, \alpha_{2}$, it is 
quite easy to obtain that in this case chaos synchronization is not 
realizable by nonreplica approach. This result conforms to the 
inference of the recent paper [21], whose author used the methods of 
differential geometry. \\
In conclusion, in this report I have shown that using the boundedness 
of the dynamical systems and nonreplica approach, one can make 
negative the real parts of the Lyapunov exponents without lengthy, 
cubmersome and tedious numerical and analytical calculations. Also it 
has been shown that the total Lyapunov exponents calculated for the 
cases of time- dependent and steady-state solutions to the dynamical 
systems, whose chaotic behavior has arisen out instability of fixed 
points, in general are different from each other.Although, it has 
established that due to the presence of many arbitrary constants in 
the response system of equations within the nonreplica approach it is 
quite  possible that in some cases these two set of Lyapunov exponents 
will be identical.\\
As an example chaos synchronization in the classical Lorenz model and 
one of R\"ossler models is investigated. Generalization of some 
features of chaos synchronization for high dimensional systems with 
some form of nonlinearities is also discussed.\\

\end{document}